\pdfoutput=1 
\documentclass[aps,pra, amsmath,superscriptaddress, longbibliography,twocolumn, showkeys, floatfix, 10pt]{revtex4-2}
\usepackage[pdftex]{graphicx}
\usepackage{bm}
\usepackage{setspace}
\usepackage{indentfirst}
\usepackage{hyperref}  
\usepackage{makecell}
\usepackage{xr-hyper}
\begin{document} 
\title{High-Temperature Quantum Emission from Covalently Functionalized van der Waals Heterostructures}

\author{S. Carin Gavin}
\affiliation{Department of Physics and Astronomy, Northwestern University, Evanston, IL 60208, USA}

\author{Hsun-Jen Chuang}
\affiliation{Materials Science and Technology Division, United States Naval Research Laboratory, Washington D.C. 20375, USA}

\author{Anushka Dasgupta}
\affiliation{Department of Materials Science and Engineering, Northwestern University, Evanston, IL 60208, USA}

\author{Moumita Kar}
\affiliation{Department of Chemistry, Northwestern University, Evanston, IL 60208, USA}

\author{Kathleen M. McCreary}
\affiliation{Materials Science and Technology Division, United States Naval Research Laboratory, Washington D.C. 20375, USA}

\author{Sung-Joon Lee}
\affiliation{Materials Science and Technology Division, United States Naval Research Laboratory, Washington D.C. 20375, USA}

\author{M. Iqbal Bakti Utama}
\affiliation{Department of Materials Science and Engineering, Northwestern University, Evanston, IL 60208, USA}

\author{Xiangzhi Li}
\affiliation{Department of Physics, Stevens Institute of Technology, Hoboken, NJ 07030, USA}
\affiliation{Center for Quantum Science and Engineering, Stevens Institute of Technology, Hoboken, New Jersey 07030, USA}

\author{George C. Schatz} 
\affiliation{Department of Chemistry, Northwestern University, Evanston, IL 60208, USA}

\author{Tobin J. Marks}
\affiliation{Department of Materials Science and Engineering, Northwestern University, Evanston, IL 60208, USA}
\affiliation{Department of Chemistry, Northwestern University, Evanston, IL 60208, USA}
\affiliation{Department of Chemical and Biological Engineering, Northwestern University, Evanston, IL 60208, USA}
\affiliation{Materials Research Center, Northwestern University, Evanston, IL 60208, USA}

\author{Mark C. Hersam}
\affiliation{Department of Materials Science and Engineering, Northwestern University, Evanston, IL 60208, USA}
\affiliation{Department of Chemistry, Northwestern University, Evanston, IL 60208, USA}
\affiliation{Materials Research Center, Northwestern University, Evanston, IL 60208, USA}
\affiliation{Department of Electrical and Computer Engineering, Northwestern University, Evanston, IL 60208, USA}

\author{Berend T. Jonker}
\affiliation{Materials Science and Technology Division, United States Naval Research Laboratory, Washington D.C. 20375, USA}

\author{Nathaniel P. Stern}
\email[]{n-stern@northwestern.edu}
\affiliation{Department of Physics and Astronomy, Northwestern University, Evanston, IL 60208, USA}

\date{\today}


\begin{abstract}

\vspace{3em}

Two-dimensional (2D) transition metal dichalcogenides (TMDs) are attractive nanomaterials for quantum information applications due to single photon emission (SPE) from atomic defects, primarily tungsten diselenide ($\text{WSe}_{\text{2}}$) monolayers. Defect and strain engineering techniques have been developed to yield high purity, deterministically positioned SPE in $\text{WSe}_{\text{2}}$. However, a major challenge in application of these techniques is the low temperature required to observe defect-bound TMD exciton emission, typically limiting SPE to $T<30$~K. SPE at higher temperatures either loses purity or requires integration into complex devices such as optical cavities. Here, 2D heterostructure engineering and molecular functionalization are combined to achieve high purity ($>$90\%) SPE in strained $\text{WSe}_{\text{2}}$ persisting to over $T=90$~K. Covalent diazonium functionalization of graphite in layered $\text{WSe}_{\text{2}}$/graphite heterostructures maintains high purity up to $T=90$~K and single-photon source integrity up to $T=115$~K. This method preserves the best qualities of SPE from $\text{WSe}_{\text{2}}$ while increasing working temperature to more than three times the typical range. This work demonstrates the versatility of surface functionalization and heterostructure design to synergistically improve the properties of quantum emission and offers new insights into the phenomenon of SPE from 2D materials.

\vspace*{3em}
\end{abstract}

\maketitle

\mbox{ }

\clearpage

\section{Introduction}

Single photon emission (SPE) from solid-state sources is valuable for quantum information applications because of its scalability and integration into optoelectronic, photonic, and quantum architectures. Various solid state SPE platforms, such as carbon nanotubes, embedded quantum dots, and crystalline defects, each bring distinct advantages and drawbacks in terms of desirable SPE characteristics like purity, brightness, indistinguishability, deterministic creation, and high working temperatures \cite{SPEsummary}. Despite this panoply, a universally ideal source is not evident, and the specific properties of diverse materials provide opportunities to explore distinct physical quantum systems and optimize SPE for different applications. For example, SPE from atomic-scale defects in two-dimensional (2D) van der Waals materials including transition metal dichalcogenides (TMDs) and hexagonal boron nitride leverages unique advantages of 2D materials to quantum optical phenomena, such as the customization of layered heterostructures and the accessibility of electrons in a 2D surface.

Within the TMD class of materials, monolayer tungsten diselenide ($\text{WSe}_{\text{2}}$) is a uniquely prolific host of SPE, with the distinct advantage of deterministically placing emitters using mechanical strain. After early investigation of this phenomenon showed that SPE from $\text{WSe}_{\text{2}}$ monolayers was detected at edges or folds \cite{tonndorf, he2015single, shang2015observation, srivastava2015optically}, a myriad of techniques have been developed to harness strain and create bright on-demand emitters \cite{nanopillar1,nanoindent,stevens2022enhancing,nanoantenna,opticalSPE}. In addition to strain engineering, techniques have developed to improve other qualities of TMD SPE such as line width and purity for better integration into functional quantum systems \cite{abramov2023photoluminescence, chemomech,graphite,SPEsummary}. 
Despite these advances in material control, viable SPE from TMDs remains challenging. The spectral energy of emitters is largely uncontrolled, spectral crowding from strain-activated defect emission can adversely affect photon purity, and the observation of SPE is typically restricted to low cryogenic temperatures. 
The typical atomic defects native to $\text{WSe}_{\text{2}}$ create defect-bound transitions with very low binding energies, meaning that SPE is generally limited to less than $T = 30$ K \cite{fang2019quick,aghajanian2023optical,tonndorf}, with the highest purity (over 90\%) restricted to less than $T = 15$ K \cite{koperski2015single, aharonovich2016solid,abramov2023photoluminescence}. SPE that can be sustained to higher temperatures either lose purity \cite{rosenberger2019quantum} or require complex fabrication such as integration into optical cavities \cite{luo2019single} or defect engineering by electron beam irradiation to create defect states deeper within the band gap \cite{irradiation}. 
Although strain engineering in TMDs has been robustly explored for controlling SPE, the 2D nature of monolayer TMDs provides additional avenues for optimization that have only recently been exploited. At low temperatures, layering graphite onto strained $\text{WSe}_{\text{2}}$ drastically improves SPE purity by quenching nearly all free and bound exciton emission \cite{graphite}. Similarly, chemical functionalization, already established as a versatile tool for the modification of nanomaterials \cite{georgakilas2012functionalization,kuila2012chemical,sinnott2002chemical,ryder2016chemically, carbenes}, can exploit the accessible surface of 2D materials to improve spectral isolation and purity of SPE in strained $\text{WSe}_{\text{2}}$ \cite{chemomech}. The interactions and bonding involved in the heterostructures and chemical functionalization are distinct, suggesting that these approaches could be synergistic, but the combination of these material approaches for modifying SPE has not yet been investigated.
In this work, we harness these two powerful tools for manipulating 2D materials to enhance SPE properties: heterostructure engineering and surface modification. The layering of graphite on nano-indented $\text{WSe}_{\text{2}}$ creates high purity ($>$90\%) SPE by suppressing competing radiative emission channels through ultra-fast charge transfer \cite{graphite}. Functionalization of this heterostructure with nitrobenzene diazonium tetrafluoroborate (4-NBD) results in covalent molecular bonds on the graphite surface, which introduces a bandgap to the semi-metal that interacts with mid-gap defect states in $\text{WSe}_{\text{2}}$. The resulting bright and pure SPE persists up to $T = 115$ K, more than triple the typical limit of similarly pure SPE in $\text{WSe}_{\text{2}}$. Mechanisms of this high-temperature SPE are discussed from both experimental and computational perspectives. These results demonstrate tailored SPE at a valuable intersection of determinism, purity, intensity, and temperature, and they present an effective approach for understanding and manipulating quantum emission in layered TMD heterostructures. 

\begin{figure*}[htbp]
    \centering
    \includegraphics[width=\textwidth]{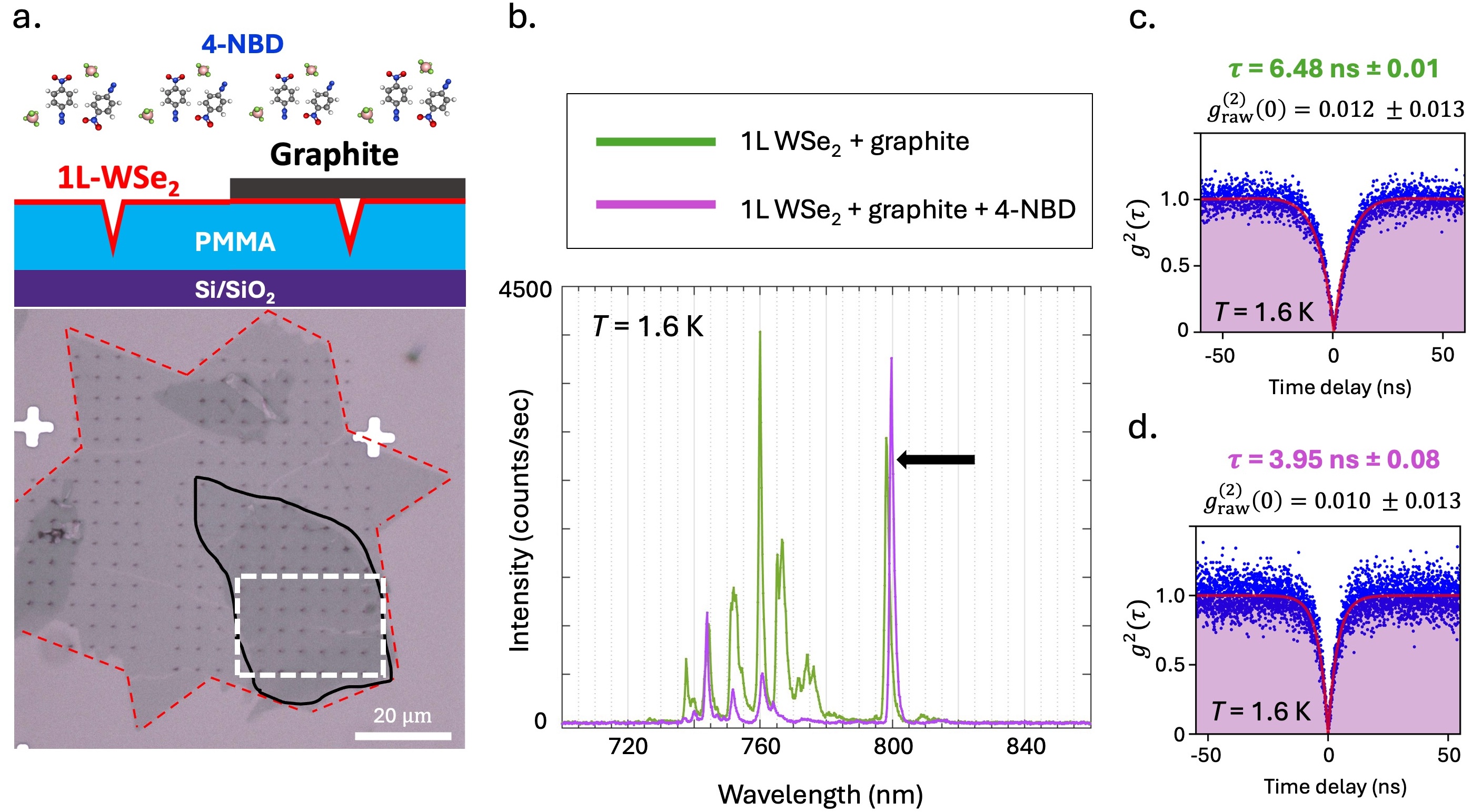}
    \caption{(a)  Optical microscope image and side-view schematic of the functionalized heterostructure. A layer of polymethyl methacrylate (PMMA) is spin-coated onto a dielectric substrate of silicon/silicon dioxide (Si/$\text{SiO}_{\text{2}}$). Monolayer (1L) $\text{WSe}_{\text{2}}$ (red) is then transferred onto PMMA and indented with atomic force microscopy (AFM). Graphite (black) is then transferred over part of this monolayer. The white dashed box outlines the area between graphite and $\text{WSe}_{\text{2}}$ that has been nano-squeegeed together. This entire heterostructure is functionalized with 4-NBD, which leaves nitrophenyl groups either covalently or non-covalently adsorbed to the surface.  (b) Spectra of the same squeegeed graphite-covered indent location before (green) and after (purple) functionalization, showing an increase in emitter intensity of the narrow emission feature at 800~nm for the same amount of laser excitation power. The laser excitation wavelength used here and throughout the work is 532~nm. (c) $g^{(2)}(\tau)$ measured for the emitter highlighted in (b) prior to functionalization. The emitter lifetime is $\tau = 6.48  \pm 0.01$~ns. (d) $g^{(2)}(\tau)$ measured for the same emitter after functionalization, where the lifetime is now $\tau=3.95\pm 0.08$~ns, a 40\% decrease in emitter lifetime.}
    \label{schematic}
\end{figure*}

\begin{figure*}[hbtp]
    \centering
    \includegraphics[width=\textwidth]{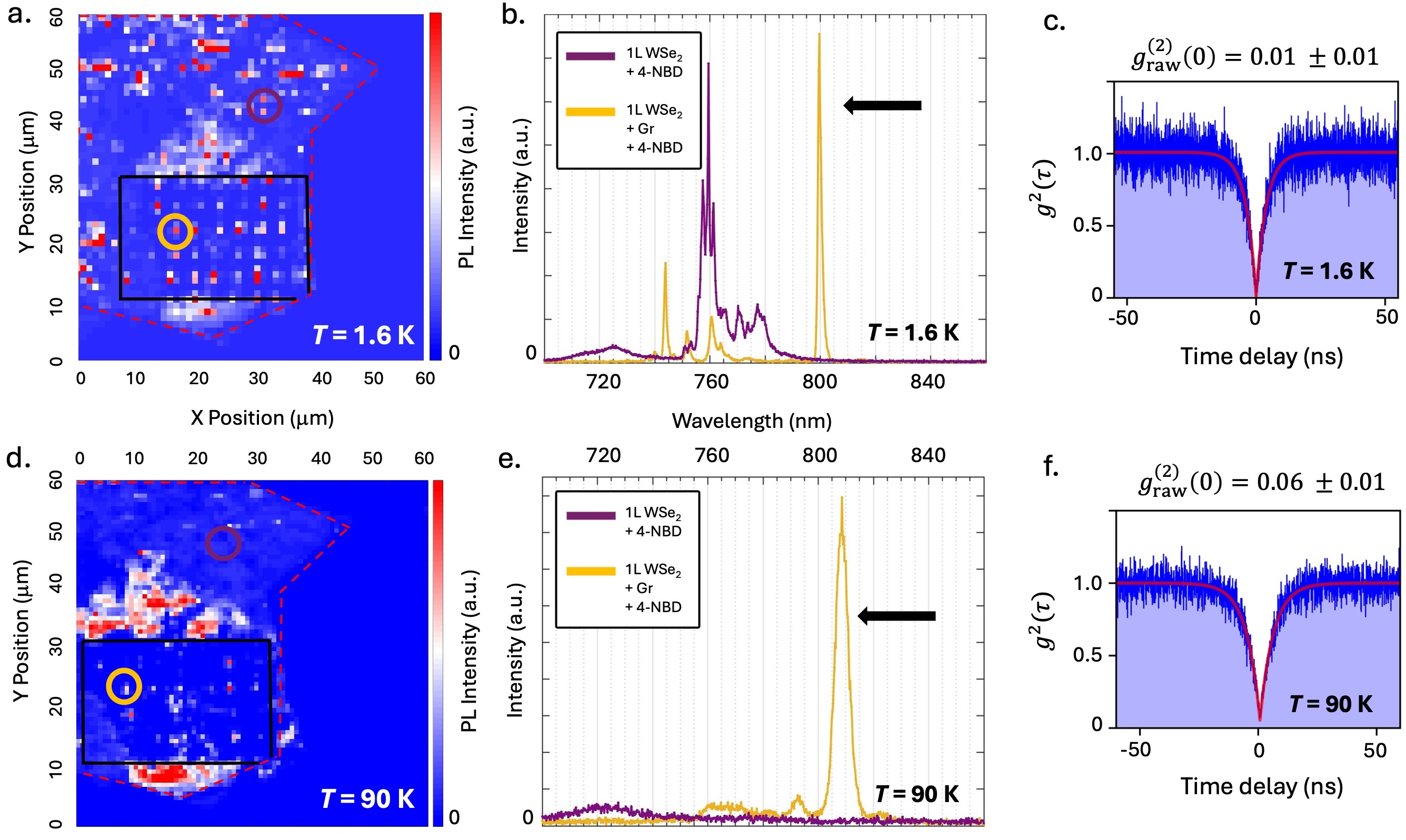}
    \caption{Results of a $\text{WSe}_{\text{2}}$/graphite heterostructure functionalized with 4-NBD; here the graphite thickness is 1.6 nm. (a) Emission intensity map of the heterostructure at $T=1.6$~K. The red dashed line outlines the $\text{WSe}_{\text{2}}$ area. The black dashed box within this area outlines the portion of graphite and $\text{WSe}_{\text{2}}$ that are squeegeed together. The relatively bright emission surrounding the squeegeed area is due to the flourescence of residue such as transfer polymer that was pushed out from between the layers \cite{rosenberger2018nano}. (b) Spectrum of two indent locations, one where just $\text{WSe}_{\text{2}}$ is functionalized with 4-NBD (purple) and one where the graphite heterostructure is functionalized (yellow). The yellow spectrum has a strong emission line around 800 nm, which is shown to be SPE at low temperature in (c). Figures (d)-(f) show the same data at $T=90$~K. In (e), the purple spectrum of the same indent as (b) shows that localized emission on the functionalized $\text{WSe}_{\text{2}}$ is gone, whereas strong emitters remain on the graphite heterostructure side (yellow). This emitter is still high purity SPE even at $T=90$~K.}
    \label{cvd-90K}
\end{figure*}

\section{Results}

\begin{figure*}[tbph]
    \centering
    \includegraphics[width=\textwidth]{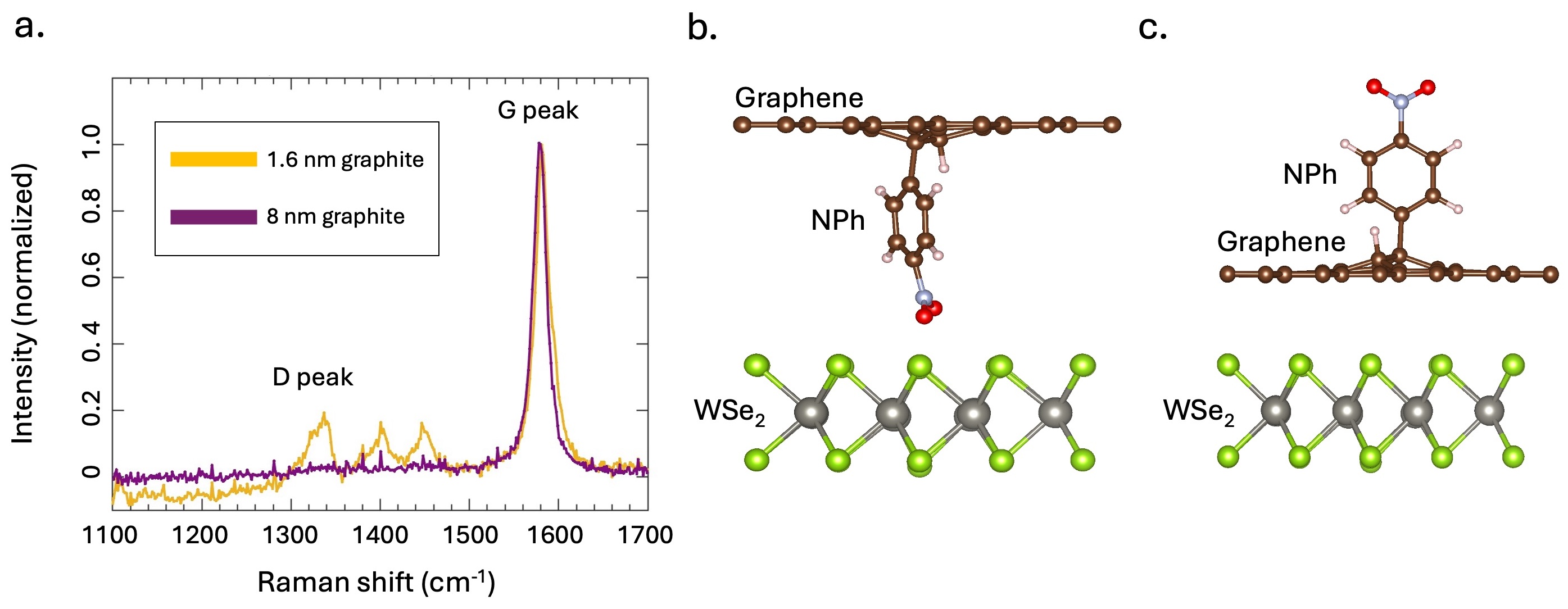}
    \caption{(a) Raman spectra of graphite after 4-NBD functionalization, one with 1.6-nm thickness (yellow) and the other with 8 nm thickness (purple). Both spectra are normalized to the graphite G peak. The D peak is detectable on the thin graphite with a resulting D/G peak ratio of about 20\%, indicating disruptions on the graphite surface due to covalent bonds. (b,c) Images of possible nitrophenyl (NPh) position in layered heterostructure configurations. In (b), NPh  is covalently attached to graphene but located between the graphene and $\text{WSe}_{\text{2}}$. In (c), NPh is covalently bonded to the top of the graphene. This configuration is more stable and  consistent with experimental findings from (a).}
    \label{graphite-raman}
\end{figure*}

\subsection{4-NBD Functionalization of Graphite/$\text{WSe}_{\text{2}}$ Heterostructures for SPE at 90 K}\label{sec:functionalizationdata}

Previous work has shown that both 4-NBD functionalization \cite{chemomech} and heterostructure engineering with graphite \cite{graphite} improve the purity of SPE from monolayers of $\text{WSe}_{\text{2}}$. These outcomes were a result of physisorption of molecules onto $\text{WSe}_{\text{2}}$ and the van der Waals interactions between $\text{WSe}_{\text{2}}$ and graphite, respectively. As such, both methods represent SPE modification based on weak electrostatic forces. Yet more powerful modification to the electronic structure comes from systems that have strong chemical, or covalent, interactions. Diazonium compounds are known to covalently functionalize carbon-based systems, such as single-walled carbon nanotubes \cite{sinnott2002chemical} and graphene/graphite \cite{kuila2012chemical, georgakilas2012functionalization,mazarei2023band, paulus2013covalent}. Based on this precedent, the functionalization dynamics in a $\text{WSe}_{\text{2}}$/graphite heterostructure should be distinct from that of $\text{WSe}_{\text{2}}$ alone. To explore this possibility and its effect on quantum emission, graphite with thickness of 1.6~nm, or about 5 layers, is transferred onto monolayer $\text{WSe}_{\text{2}}$ grown by chemical vapor deposition (CVD) (Figure \ref{schematic}a). The heterostructures are prepared following the method of Ref. \cite{graphite}, with strain introduced using a nanoindentation array and the layers ``nanosqueegeed" to remove impurities between them and improve contact quality, according to the precedent established by Rosenberger \textit{et al.} \cite{rosenberger2019quantum,rosenberger2018nano}. This hybrid method consistently produces large monolayer areas and a reliably high yield of SPE from the strain array \cite{rosenberger2019quantum}, both of which are valuable for the scalability of quantum emission since monolayer area achieved by micromechanical exfoliation is variable and much smaller on average \cite{novoselov2005two,li2022recent}. The heterostructure is then functionalized by being submerged in an aqueous solution of 4-NBD. This process dopes the materials present since 4-NBD is electrophilic, and nitrophenyl (NPh) radicals form as a result. These radicals may either bond to the surface covalently or self-react, forming chains of multiple nitrophenyls (oligomers), as described in previous work \cite{chemomech}. Further details of sample fabrication are found in Section \ref{sec:methods}.

Figure \ref{schematic}b shows the effect of functionalization on SPE from the $\text{WSe}_{\text{2}}$/graphite heterostructure at low temperature.  For the same graphite-covered indent location at $T=1.6$~K as measured with the same amount of laser excitation power, many higher-energy (below $\sim$780 nm) localized emitters are quenched by 4-NBD, whereas the intensity of the emitter at 800~nm is increased by functionalization. Figures \ref{schematic}c,d show $g^{(2)}(\tau)$ measurements for this emitter, demonstrating that it is SPE of high purity above 90\%. The lifetime of SPE extracted from the correlation relaxation from $\tau = 0$ is reduced by functionalization from $\tau=6.48$~ns with graphite alone to $\tau=3.95$~ns after functionalization. The impact of 4-NBD functionalization on the temporal dynamics of SPE in $\text{WSe}_{\text{2}}$/graphite heterostructures is a strong indication that this structure has changed the mechanisms underlying the phenomenon. 

To highlight that both 4-NBD and graphite contribute to changed dynamics, Figure \ref{cvd-90K} compares the low and high temperature performance of SPE from the functionalized heterostructure to the plain $\text{WSe}_{\text{2}}$ functionalized with 4-NBD. At low temperatures ($T=1.6$~K), the results are reliable: sharp, localized emission is observed from indent locations (Figures ~\ref{cvd-90K}a-c). The SPE spectra from indented $\text{WSe}_{\text{2}}$ (dark purple) are consistent with previous reports of 4-NBD functionalization of strained $\text{WSe}_{\text{2}}$, showing a red-shifted exciton and suppression of most excitonic emission \cite{chemomech}. The graphite-covered indent (the same as that shown in Figure \ref{schematic}b) exhibits relatively more quenching of excitonic features, and fewer isolated emission lines remain, but the high purity SPE remains at 800~nm (yellow) (Fig.\ref{cvd-90K}c). 

Although the results for $T=1.6$~ K are in line with previous expectations, the combination of graphite and 4-NBD surprisingly sustains the high-purity SPE to much higher temperatures. Figures \ref{cvd-90K} (d)-(f) show the properties of the same indent locations (circled in purple and yellow) at $T=90$~K. At this temperature, sharp emission from the plain $\text{WSe}_{\text{2}}$ indent has disappeared, in accordance with the typical temperature restrictions of defect-bound emission in TMDs. In contrast, intense sharp features persist on the graphite-covered indents. These features are once again shown to be SPE with purity over 90\%. In addition to high purity, the peak intensity remains high at $T=90$~K with a maximum value of approximately 800 counts/sec, resulting in an excellent signal-to-noise ratio in the measurements $g^{(2)}(\tau)$ (Figure \ref{cvd-90K}f). Although this is the cut-off temperature for purity over 90\%, the emitter remains viable with $g^{(2)}(0)< 0.5$ up to $T=115$~K and detectable in the spectrum to $T=150$~K, on par with other methods producing higher temp SPE in $\text{WSe}_{\text{2}}$ \cite{luo2019single,irradiation}. A comparison of temperature, purity, and emitter lifetime is shown in Figure S\ref{supp-g2-temp}, and the full temperature evolution is presented in Figure \ref{temp}c in Section \ref{sec:discussion}.

Not all SPE observed on the functionalized heterostructure at low temperatures remains up to $T=90$~K. Rather, localized emission features that persist to higher temperatures lie within a narrower wavelength range than those visible at low temperatures. However, the yield of these emitters across the strain array is high; the ratio of candidate SPE at low temperatures which persist to higher temperatures is about 75\%, discussed in Figure S\ref{supp-histo}. Figure S\ref{supp-bonus 90K} details some of these additional emission spectra and correlation measurements confirming SPE from graphite-covered indents present in Figure \ref{cvd-90K}d. The high-intensity emission peaks at this temperature are centered between 780-820~nm, whereas for $T <10$~K SPE are commonly anywhere between 720-820~nm \cite{srivastava2015optically,koperski2015single, nanopillar1,chemomech}. In other words, lower-energy SPE transitions selectively last to higher temperatures, while higher-energy SPE transitions are quenched at the typical temperature scales. This behavior is similar to the effect of 4-NBD functionalization at low temperatures, where higher energy emission is suppressed after functionalization, but the lower energy SPE intensity is increased~(Fig. \ref{schematic}b).   

\subsection{Covalent Functionalization of Graphite}

Following the novel results discussed above, Raman spectroscopy was used to assess the nature of the 4-NBD functionalization and its interaction with the 2D layers in the heterostructure. Pristine graphite has a characteristic Raman feature centered around 1580 cm$^{\text{-1}}$ called the `G peak' \cite{ferrari2007raman}; when there are disturbances in the carbon bonds on the surface as a result of covalent functionalization, a new Raman mode appears around 1350 cm$^{\text{-1}}$ , known as the `D peak' \cite{georgakilas2012functionalization}. The ratio of these two peaks (D/G) characterizes the degree of covalent functionalization on the graphite surface. Figure \ref{graphite-raman}a shows the Raman spectra of two different functionalized heterostructures made with different thicknesses of graphite, one in which graphite is approximately 8 nm thick, and the other in which our high temperature SPE were measured, where graphite is 1.6 nm thick. Thin graphite should have a higher degree of covalent functionalization than thick graphite overall given the lower density of states \cite{georgakilas2012functionalization}. Here, the D peak is not detected on very thick graphite. However, for the heterostructure used in this study of thinner graphite, the D/G peak ratio is approximately $20\%$, referenced with the primary D peak at 1350 cm$^{\text{-1}}$. Two additional Raman peaks are observed around 1400 and 1450 cm$^{\text{-1}}$, which can indicate other types of lattice disruptions. This confirms that our specific heterostructure has covalent bonds between NPh and graphite from the functionalization process, which facilitates the observation of SPE up to $\sim90~K$. Figures \ref{graphite-raman}b,c show schematics of two possible heterostructure arrangements: one in which the nitrophenyl group is mainly between $\text{WSe}_{\text{2}}$ and graphene and one in which it is mainly on top of graphene. To support the interpretation that 4-NBD preferentially functionalizes the top surface of graphite, calculations were performed to find the most energetically stable chemical configuration of these elements. The total energy of surface functionalization when NPh is on top of graphene is lower by 2.62~eV compared to between layers, indicating that it is more stable. This calculation combined with the Raman spectra reveal that the arrangement in Figure \ref{graphite-raman}c is the likely configuration of the heterostructure.

\begin{figure}[tbph]
    \centering
    \includegraphics[width=\columnwidth]{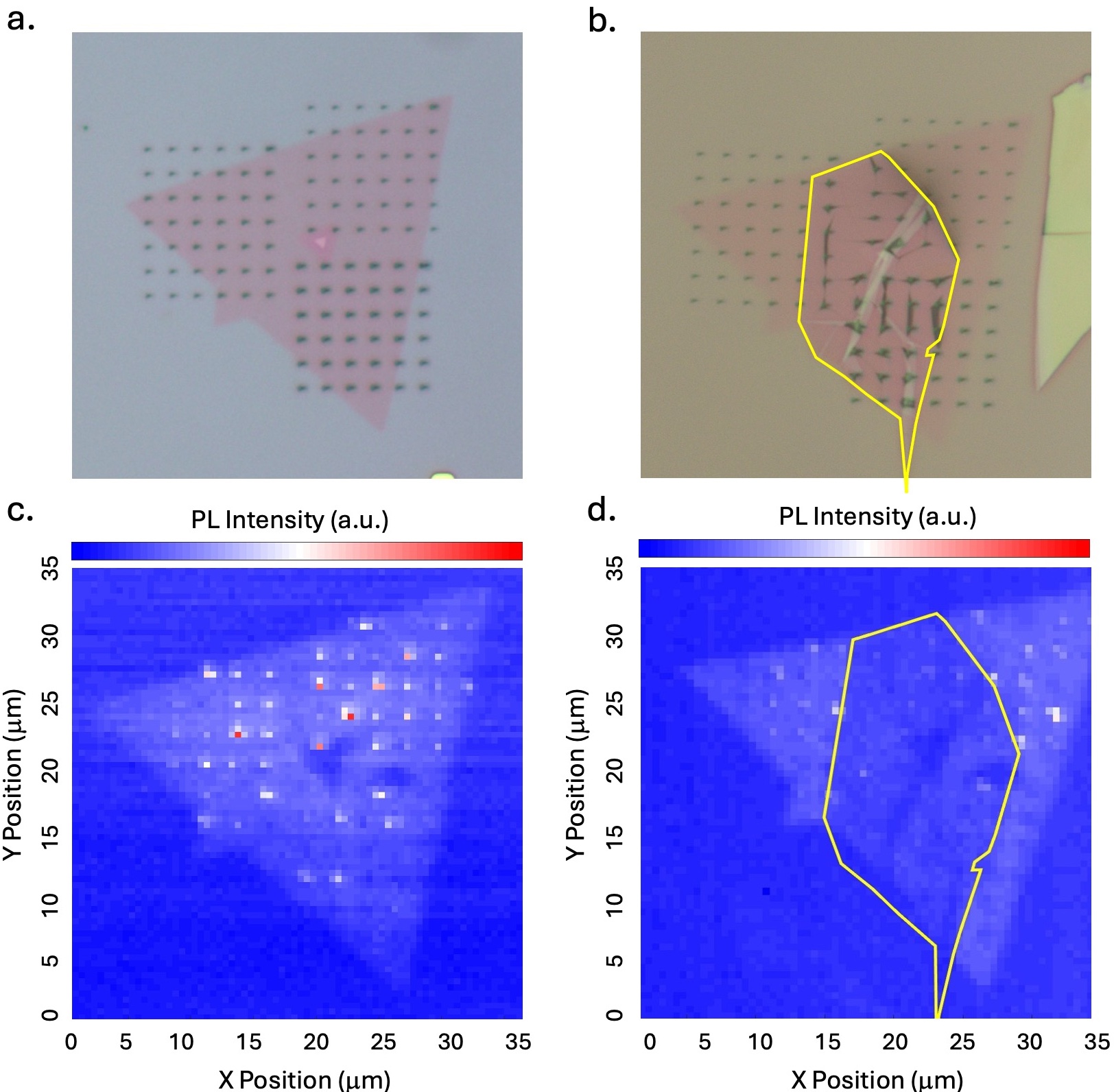}
    \caption{Results of a heterostructure with graphite transferred directly on top of functionalized monolayer $\text{WSe}_{\text{2}}$. (a) OM of a functionalized monolayer. (b) OM of this same monolayer with graphite (outlined in yellow) transferred on top. (c) Intensity map showing maximum peak intensity of the functionalized monolayer without graphite. (d) Intensity map of the functionalized monolayer with graphite (outlined in yellow) transferred on top. Here, the graphite suppresses nearly all emission from both flat and strained areas. Localized emission from indents is very weak and disappears by $\sim10K$.}
    \label{control test}
\end{figure}

\begin{figure*}[tbph]
    \centering
    \includegraphics[width=\textwidth]{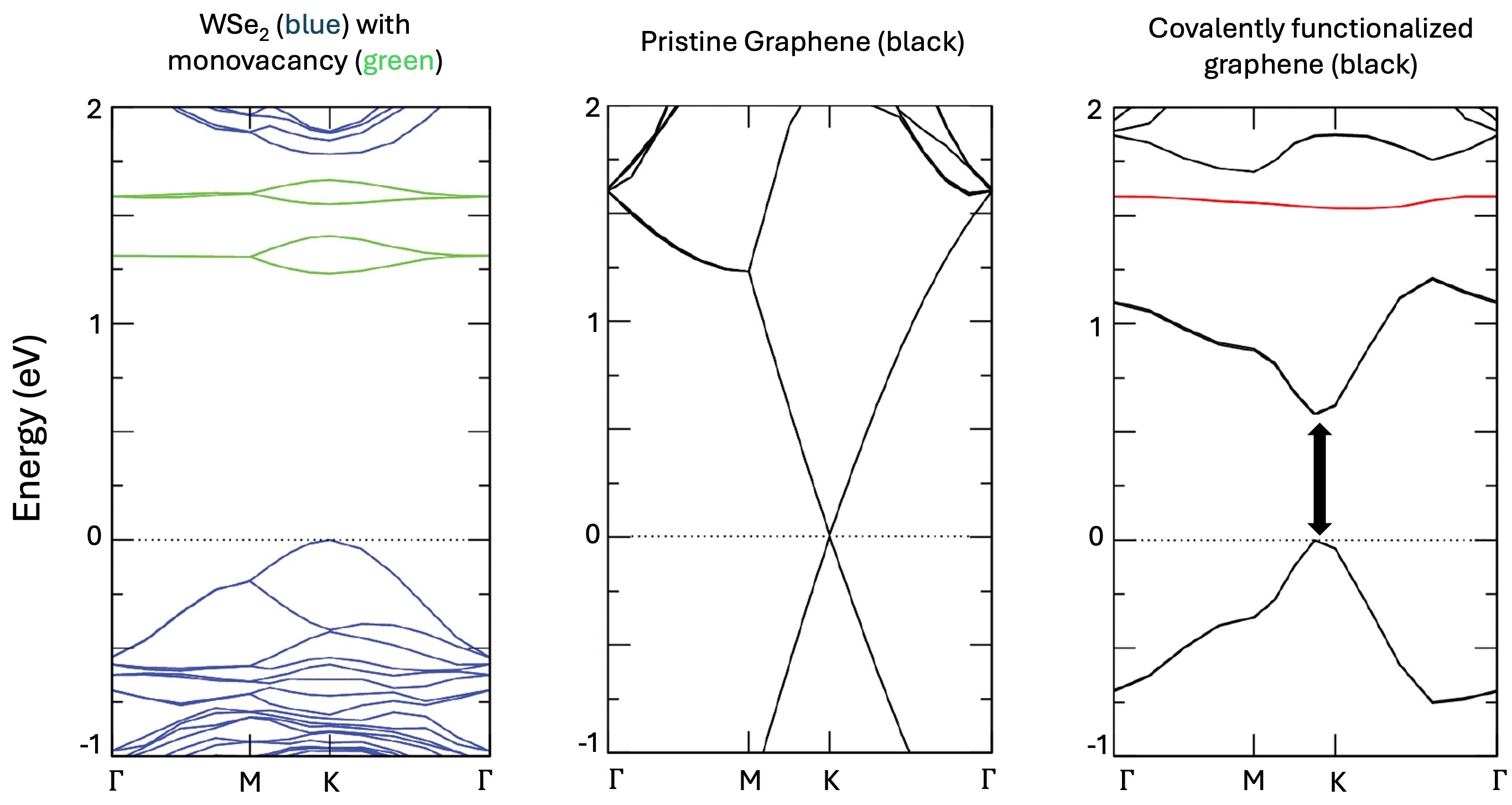}
    \caption{The electronic band structure of monolayer $\text{WSe}_{\text{2}}$ (blue) with a single selenium vacancy (green), pristine graphene (black), and covalently functionalized graphene (black) with an additional energy level introduced by nitrophenyl (red). The valence band maximum is shifted to zero and is shown by the black dashed line.}
    \label{bandstructure}
\end{figure*}

To definitively test whether the improved purity and elevated temperatures of Section \ref{sec:functionalizationdata} are due to the covalent interaction of diazonium on graphite specifically, two control tests were performed. The precedent of diazonum functionalization of $\text{WSe}_{\text{2}}$ was established with exfoliated monolayers. Since our results were obtained with CVD $\text{WSe}_{\text{2}}$, which has a higher defect density on average \cite{defectdensity, aryeetey2020quantification}, the 4-NBD functionalization of the latter is tested alone to ensure that the effects are consistent between materials. Indeed, the modification is consistent with that observed with exfoliated material; emission from flat regions of the monolayer is suppressed, but SPE on the strain array remains (Figures S\ref{supp-cvd-diazonium} and S\ref{supp-cvd-g2}). This further demonstrates that 4-NBD functionalization is a versatile method, effectively enhancing spectral isolation and purity in both exfoliated and CVD monolayers, using both nanopillars and nano-indented strain arrays. However, these properties cannot be sustained beyond the typical temperature scales, even with the varied defect environment of CVD monlayers. As another test, the same $\text{WSe}_{\text{2}}$ had graphite transferred onto it after functionalization was performed, and thus oligomerization had already occurred. As seen in Figure \ref{control test}, graphite suppresses the emission in both the flat and strained parts of the flake, as it does on $\text{WSe}_{\text{2}}$ alone \cite{graphite}. The enhanced intensity and elevated temperatures cannot be replicated when the non-covalent functionalization process occurs with $\text{WSe}_{\text{2}}$ rather than a covalent functionalization with graphite (Fig. \ref{control test}c,d). 
This set of characterization techniques illustrates a clear picture of how the 4-NBD functionalization occurs in the experimental heterostructure. Thin graphite ($\sim 1.6$~nm) is flexible and in close contact with $\text{WSe}_{\text{2}}$ after the nano-squeegee process. The nitrophenyl groups covalently functionalize the top graphene surface, evidenced by stability calculations and Raman spectroscopy showing new covalent bonds. Physisorption of oligomers to $\text{WSe}_{\text{2}}$ or graphite as a cause is ruled out, demonstrating that high-purity and elevated temperature observations can only be produced when the graphite surface is covalently functionalized. 

\subsection{Modeling}

Chemical functionalization with aryl diazonium groups is an established tool to tune 2D materials, including graphite \cite{park2013covalent, mazarei2023band, huang2013diazonium, paulus2013covalent}, but this modified graphite has yet to be leveraged for enhancing the properties of quantum emission. With experimental and computational evidence that 4-NBD covalently functionalizes the top graphite surface of our heterostructure, density functional theory (DFT) calculations were performed to demonstrate how the altered graphite affects the electronic band structure and subsequent SPE properties. Figure \ref{bandstructure} shows the band structure of defective monolayer $\text{WSe}_{\text{2}}$, pristine graphene, and covalently functionalized graphene. It is well established that 1L $\text{WSe}_{\text{2}}$ is a direct bandgap semiconductor \cite{sun2016indirect, yang2016manipulation}, with midgap defect states introduced by common defects, such as selenium vacancies. Defect-bound transitions through the resulting energy levels are proposed as the source of SPE \cite{gavin2024role,irradiation,chemomech,cui2018effect,kapuscinski2021rydberg,hernandez2022strain, zheng2019point,zhang2017defect}. Pristine graphene exhibits its characteristic Dirac cone at the K point \cite{partoens2006graphene, castro2009electronic}. However, when new chemical bonds are introduced on the graphene surface, a bandgap opens in the graphene energy levels such that it is no longer a semi-metal structure. This new ``semiconductor graphene'' has a gap magnitude consistent with previous reports \cite{huang2013diazonium} and resides within the $\text{WSe}_{\text{2}}$ band gap, which means that the graphene levels interact with the defect states that produce SPE. This has critical implications for the resulting behavior between layers. First, the fact that graphite is no longer a semi-metal means that carrier mobility has been reduced. Higher degrees of chemisorption reduce the carrier mobility of both electrons and holes in graphene \cite{paulus2013covalent,huang2013diazonium}, which localizes carriers at these sites. Furthermore,  a viable optical transition from midgap energy levels above and below the Fermi energy has been created by functionalized graphene. Direct, defect-centered transitions have been proposed as the source of SPE \cite{gavin2024role}, and since such transitions are also created by specific defect configurations in $\text{WSe}_{\text{2}}$, it is possible for both to interact to create more strongly bound transitions than typically exist in $\text{WSe}_{\text{2}}$ alone. Since low binding energy of defect-bound transitions in $\text{WSe}_{\text{2}}$ is their typical weakness, the interaction with a `defect-centered graphene transition' can localize charge and optically recombine with $\text{WSe}_{\text{2}}$ defect levels, or even resonate with a specific transition.

Here, it is important to note that many factors affect the precise magnitude of both defect transition energy gaps and the functionalized graphite bandgap. For $\text{WSe}_{\text{2}}$, we must first consider the limits of DFT in predicting the magnitude of energy gaps beyond the neutral exciton, particularly for fine structure and defect-bound emission. DFT demonstrates how energy levels evolve in a crystal given specific conditions but does not predict the spectral energy of SPE. Furthermore, since the exact defects and mechanisms responsible for SPE are still largely debated \cite{gavin2024role,irradiation,chemomech,cui2018effect,kapuscinski2021rydberg,hernandez2022strain, zheng2019point,zhang2017defect}, each model will produce a different alteration to the bandgap. Next, it is critical to consider the role of mechanical strain. Strain is an experimental necessity for observing SPE in this material, as previously discussed. But strain also alters other optoelectronic properties of semiconductors and can modulate the bandgap of $\text{WSe}_{\text{2}}$, generally red-shifting its emission \cite{yang2016manipulation,islam2024strain, qi2023recent,blundo2021strain, aslan2018strain}. The properties of SPE are very sensitive to microscopic strain variations. As such, the precise energy gap of any single SPE transition in a given strain environment is very difficult to predict. 

Another factor to consider is the degree of covalent functionalization present on the graphene or graphite surface. In this work, the graphene model has only one out of every 50 carbon atoms covalently attached to a nitrophenyl group, and this is enough to open a small bandgap. If the concentration of nitrophenyl bonds is locally higher and the graphene degeneracy is broken further, the magnitude of the bandgap will increase. In this regard, strain plays another important role. The density of covalent functionalization of 4-NBD on graphite depends largely on two factors: material thickness and amount of strain \cite{paulus2013covalent}. As demonstrated by our Raman spectroscopy, the degree of chemisorption between diazonium and graphite increases with decreasing layer number (Figure \ref{graphite-raman}a). Furthermore, strain on the graphite surface can help locally increase the density of covalent bonds \cite{paulus2013covalent}. Although the graphite in these samples is not indented itself, it does not lie completely flat on the indented $\text{WSe}_{\text{2}}$. Where the schematic in Figure \ref{schematic}a shows a simplified device architecture, an accurate nano-indent profile is shown in Figure S\ref{supp-profile}. The indents are asymmetric and have ``shoulders'' instead of a flat opening. When few-layer graphite is transferred onto this profile, it conforms to and adopts this curvature as well. This, in turn, can increase the degree of covalent functionalizaton on the graphite at that site.

Given all these considerations, our model represents the underlying principles at work in the system and encapsulates the key features of the sample: defects create midgap states in $\text{WSe}_{\text{2}}$, and covalently functionalizing graphene opens a small bandgap within that of pristine $\text{WSe}_{\text{2}}$. The magnitudes of each can vary, leading to complex interactions or even potential resonance between energy levels, and such interacting midgap transitions increase the binding energy of single-electron transitions.

\section{Discussion}\label{sec:discussion}

\begin{figure}
    \centering
    \includegraphics[width=\columnwidth]{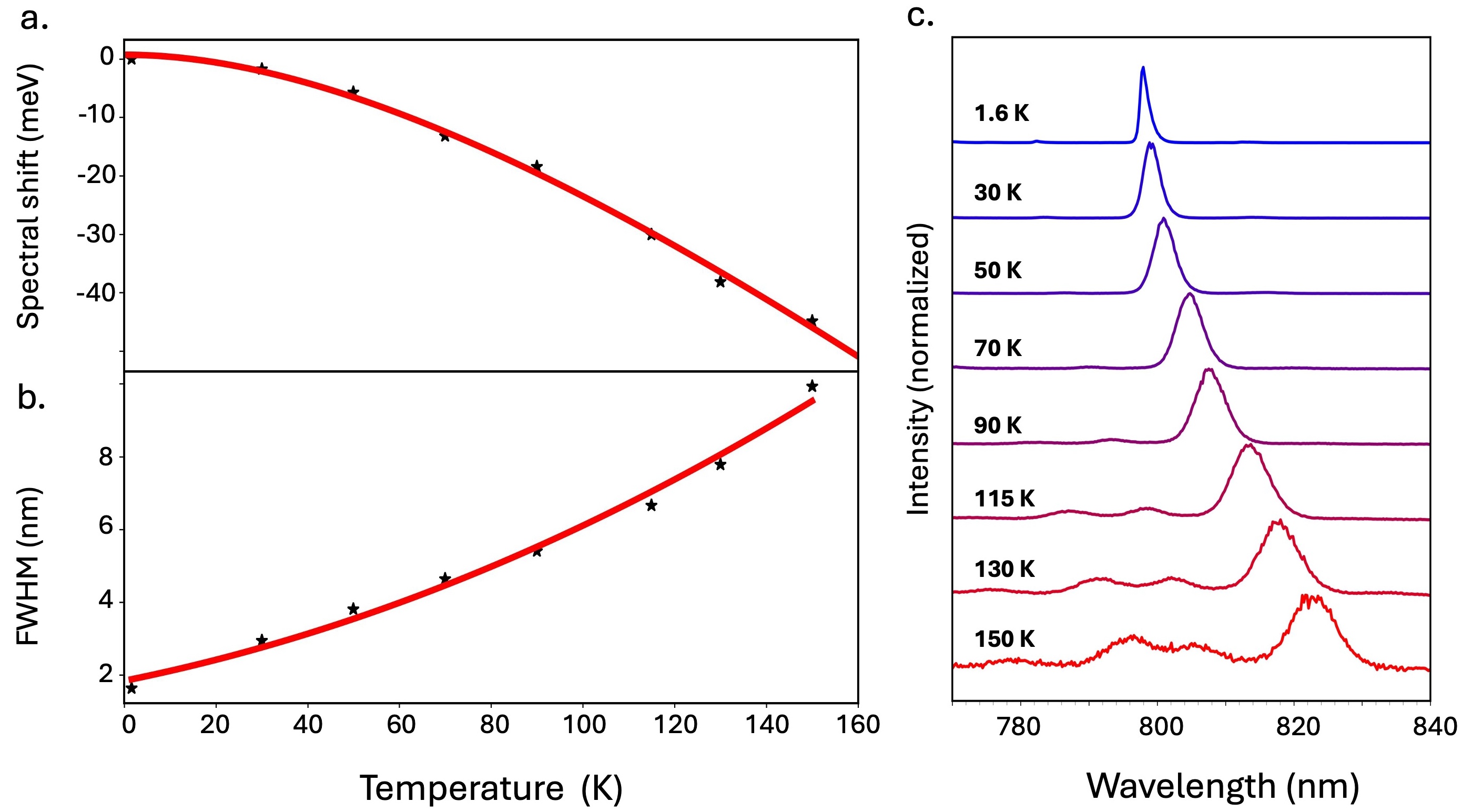}
    \caption{(a) Spectral shift of SPE PL peak center with temperature. Data points are shown as black stars, and the Varshni fit line is shown in red (precedent for this fit established by Parto \textit{et al.} \cite{irradiation}). (b) Evolution of SPE linewidth as a function of temperature for the same source. Data points are shown as black stars, and the quadratic fit line is shown in red. (c)  PL spectra of the SPE with increasing temperature (spectra offset for clarity).}
    \label{temp}
\end{figure}

The system presented here is complex, with crystalline defects, vdW material layer interactions, and covalently bonded molecules. From this combination, high-purity SPE at elevated temperatures was achieved. Given this complexity, it is helpful to discuss the results in the context of other reports of elevated SPE temperatures in 2D $\text{WSe}_{\text{2}}$. Parto \textit{et al.} demonstrated SPE at $\sim 150$~K by irradiating $\text{WSe}_{\text{2}}$ to induce defects that can have binding energies higher than normal \cite{irradiation}. The SPE observed here have some quantifiable similarities, such as redshifting with increased temperature (Figures \ref{temp}a,c) and broadening linewidth (Figure \ref{temp}b). Yet our approach operates on the distinct premise of using intrinsic defects in their typical densities, indicating that the mechanisms of these results are different. In another example, Luo \textit{et al.} integrated $\text{WSe}_{\text{2}}$ into plasmonic nanocavities \cite{luo2019single} to observe SPE up to 160 K. Strongly coupled systems such as optical cavities enhance light-matter interactions \cite{luo2024strong}, with the Purcell-enhanced ratio of radiative to non-radiative recombination in nanocavity-coupled $\text{WSe}_{\text{2}}$ leading to higher accessible temperatures for SPE. The explanation in \cite{luo2019single} based on enhanced quantum yield provides a possible interpretation of our work. As noted, the graphite bandgap can be resonant with $\text{WSe}_{\text{2}}$ midgap states under the correct defect and strain conditions.

However, a notable difference in our work compared to the nanocavity-coupled SPE is the high photon purity at elevated temperature. The SPE presented in Figures \ref{cvd-90K} and S\ref{supp-bonus 90K} exhibit spectral broadening compared to their low-temperature counterparts, with an average full width half maximum (FWHM) of $\sim$5~nm compared to $<$2~nm, respectively (Figure \ref{temp}b). Generally, such broadening due to effects like phonon scattering and exciton-phonon coupling \cite{he2016phonon, antonius2022theory} causes the purity of SPE to decrease when temperatures increase. However, the emitters presented here exhibit very high purity at $T=90$~K even with spectral broadening, which means that scattering and other competing recombination mechanisms are not being introduced with increased temperature, as they are in other systems. To our knowledge, purity over 90\% has not yet been reported in SPE beyond $\sim10$~K \cite{irradiation, luo2019single, stevens2022enhancing}. 

As temperature increases, the non-radiative relaxation rate also increases. In these prior realizations, the quantum yield decreases. For the purity of $g^{(2)}(0)$ to decrease, additional radiative pathways must be excited beyond the relevant single-photon transition. Broadening of the emission from non-radiative processes at elevated temperature does not necessarily pollute the photon purity, although these are correlated in prior experiments. In our case, the combination of the functionalization and the graphite heterostructure appears to sufficiently quench undesired radiative emission, even while allowing for broadening from non-radiative emission. 

The experimental features of broadened emission and high purity also correlate with expectations of collective Dicke enhancement from an ensemble of transitions \cite{dicke1954coherence, breeze2017room,tufarelli2021single}. Dicke enhancement, or Dicke superradiance, can be observed for single photon sources such as quantum dots \cite{scheibner2007superradiance,tighineanu2016single,zhu2024single,tufarelli2021single}. It is marked by properties such as enhanced emitter intensity (and the corresponding reduction of emitter lifetime) \cite{tighineanu2016single}. Although these are observed and the defect density is certainly high enough for emitter separation in the crystal to be within the wavelength scale of optical coherence needed for collective effects~\cite{masson2022universality}, it is clear that Dicke enhancement does not exist in typical SPE observed in $\text{WSe}_{\text{2}}$, nor in a heterostructure with graphite as a semi-metal. Although the spatial requirement for superradiance is met by the distance between atomic defects in the monolayer $\text{WSe}_{\text{2}}$, the system lacks the energetic resonance required to facilitate collective behavior \cite{dicke1954coherence}. The elevated temperatures and high purity are more likely caused by the available transition pathways in the graphite heterostructure with 4-NBD rather than novel collective behavior of multiple defects in close proximity. 

In conclusion, we have demonstrated that diazonium functionalization is a broadly applicable and scalable method for improving the purity of SPE. More importantly, we have shown that SPE of high purity can be sustained to $T=90$~K in a functionalized $\text{WSe}_{\text{2}}$/graphite heterostructure and observed with $g^{(2)}(0)< 0.5$ up to $T=115$~K. Materials characterization, modeling, and control measurements suggest that this is caused by the covalent bonding between nitrophenyl and the graphite surface. DFT reveals that covalently attached nitrophenyl opens a bandgap in graphene that exists within the bandgap of $\text{WSe}_{\text{2}}$ and localizes charge carriers. This resulting interaction with midgap defect states from defects creates bright, pure SPE that exceeds the typical temperature constraints of those qualities by roughly 80 K. Combined with prior reports, this result adds further evidence that chemical functionalization is a viable pathway for tailoring TMD optical properties beyond purity and can alter the fundamental mechanisms of a heterostructure when interacting covalently. This molecular and structural engineering of TMD defect emission demonstrates a new framework to control quantum photon sources that harnesses the standout benefits of this class of materials. 

\section{Methods}\label{sec:methods}

\subsection{Material Growth}

For CVD-grown sample configurations, the monolayer $\text{WSe}_{\text{2}}$ was synthesized on Si/$\text{SiO}_{\text{2}}$ substrates featuring a 275 nm oxide layer, using quartz tube furnaces according to established procedures. To facilitate lateral growth, perylene-3,4,9,10-tetracarboxylic acid tetrapotassium salt molecules are deposited on the growth substrates. High-purity metal oxide and chalcogen powders are employed as precursors, while a continuous flow of ultra-high purity argon is maintained during the heating process to approximately 850 $^{\circ}$C. Upon reaching the target temperature, ultra-high purity hydrogen is introduced into the argon stream and sustained throughout the soak period and subsequent cooling to ambient temperature \cite{mccreary2016synthesis}.

\subsection{Sample Fabrication}

Two types of sample configurations were carefully fabricated on the PMMA substrate for SPE generation. For both samples types, polymethyl methacrylate (PMMA) 950A4 was spun onto a clean $\text{SiO}_{\text{2}}$ substrate and cured at 180 $^{\circ}$ C for 2-3 minutes to obtain the 400nm thickness of PMMA in the $\text{SiO}_{\text{2}}$ wafer. To obtain the first sample configuration (exfoliated 1L-$\text{WSe}_{\text{2}}$/PMMA) we directly exfoliate monolayers of $\text{WSe}_{\text{2}}$ onto PMMA/$\text{SiO}_{\text{2}}$; the $\text{WSe}_{\text{2}}$ bulk crystals were purchased from HQ Graphene. To obtain the second sample configuration (CVD-1L-$\text{WSe}_{\text{2}}$/PMMA), we use a solution-assisted removal and transfer method to place CVD-grown monolayer $\text{WSe}_{\text{2}}$ on PMMA /$\text{SiO}_{\text{2}}$. 

The AFM nano-indentation was then performed to generate the indentation array on both types of the sample configuration mentioned above, followed by the low-temperature photoluminescence to confirm the existence of localized spectral features of the $\text{WSe}_{\text{2}}$ indentation sites. Thin graphite flakes are then mechanically exfoliated onto either a PDMS patch or $\text{SiO}_{\text{2}}$ substrate, subsequently identified by optical contrast and measured using AFM to determine thickness. A graphite flake of desired thickness is then directly transferred on top of the indented-$\text{WSe}_{\text{2}}$/PMMA substrate to partially cover the indented $\text{WSe}_{\text{2}}$ array. All detailed steps of the sample fabrication can be found in previous studies \cite{rosenberger2019quantum,chuang2016low,graphite}.

\subsection{Materials Characterization and Chemical Functionalization}

Raman spectroscopy was performed under ambient and room temperature conditions using a Horiba XploRA Plus instrument and excitation with a 532-nm laser (100x objective, NA 0.9). For chemical functionalization, the diazonium salt, 4-nitrobenzenediazonium tetrafluoroborate (4-NBD), was purchased from Sigma-Aldrich (97\%), stored at $4^\circ$ C and purified by recrystallization. An aqueous 5mM solution of 4-NBD was prepared no more than 15 minutes prior to use. The $\text{WSe}_{\text{2}}$ or $\text{WSe}_{\text{2}}$-Gr samples were immersed in 5mM 4-NBD for 90 min in a glass scintillation vial that was shielded from light with aluminum foil. After immersion, the samples were rinsed with deionized (DI) water and dried with nitrogen flow.

\subsection{Low Temperature Optical Spectroscopy}

Cryogenic optical spectroscopy and photon correlation measurements were performed in an AttoDry 2100 cryostat starting at $T=1.6$ K and up to $T=150$ K. High-resolution spatial mapping was performed using Attocube piezoeletric nanopositioners in the $x-y$ plane within the cryostat. Confocal spectroscopy was performed using a 0.82 NA, $100\times$ magnification objective with a 532 nm diode laser excitation. The diffraction-limited spot size is estimated to be $D=1.22\lambda/NA \approx 0.79$~$\mu m$. Light was collected with an optical fiber and sent to a 750-mm focal length spectrometer (Andor Shamrock SR-750) with a thermoelectrically cooled CCD camera (DU420A-BEX2-DD). A Hanbury-Brown-Twiss setup was used to measure the coherence. The signals were sent to an optical fiber with a $1\times2$ fiber splitter to direct equal signals to two avalanche photodiodes (APD; PicoQuant, $\tau$-SPAD-100). To isolate light from the emitter of interest, a 10-nm FWHM bandpass filter of the appropriate center wavelength was used. The raw photon coincidence data was used to calculate the second-order intensity correlation function $g^{(2)}(\tau)$.  Background correction was initially applied to check the final $g^{(2)}(0)$ value. This correction only accounts for the background arising from dark counts on the APD from light contamination of the measurement environment. The background count rate was obtained from the APD with laser illumination blocked. For each SPE source presented in this work, the background correction did not alter the extracted $g^{(2)}(0)$ value, and therefore the raw values are reported.

\subsection{DFT Calculations}

For DFT modeling, all calculations are performed using projector-augmented wave (PAW) \cite{blochl1994projector} and generalized gradient approximation (GGA) \cite{perdew1992jp} with the Perdew, Burke, and Ernzerhof (PBE) exchange correlation functional as implemented in Vienna ab initio simulation package (VASP) \cite{kresse1996efficient}. To account for van der Waals (vdW) interactions, the DFT-D3 correction proposed by Grimme was used \cite{grimme2010consistent}. The cutoff energy for the plane wave function is 450 eV.  For geometry optimization, a $4 \times 4 \times 1$ Monkhorst-Pack k-grid \cite{monkhorst1976special} is used, and for electronic structure calculations an $8 \times 8 \times 1$ grid is used. The energy convergence criterion was set to 0.0001 eV and the whole system relaxed until the maximum forces on the atoms are lower than 0.001 eV/\AA. Electronic band structures were calculated using Heyd-Scuseria-Ernzerhos hybrid functional (HSE06) coupled with spin-orbit coupling (SOC) \cite{paier2006screened}. VASPKIT \cite{wang2021vaspkit} was used for some post-processing band structure calculations. $\text{WSe}_{\text{2}}$ was modeled with a $4\times 4 \times 1$ supercell and graphene was modeled using a $5 \times 5 \times 1$ supercell. The heterostructure consists of $\text{WSe}_{\text{2}}$ having a 4 $\times$ 4 $\times$ 1 supercell and graphene having a $5 \times 5 \times 1$ supercell to minimize the lattice mismatch. When graphene is covalently functionalized with diazonium, for charge balancing, one H atom is attached to the neighboring carbon where the covalently bonded nitrophenyl group is attached.

\section{Author Contributions}

S.C.G performed all optical spectroscopy and photon counting. A.D. performed chemical functionalization and Raman spectroscopy. M.K. executed first-principles calculations. H-J.C., S-J.L., X.L., K.M.M., and M.I.B.U. fabricated monolayer $\text{WSe}_{\text{2}}$ indentation and heterostructure samples. S.C.G. prepared the manuscript with the input of all authors. G.C.S., T.J.M., M.C.H., B.T.J., and N.P.S. supervised the project. 

\section*{Competing Interests}

The authors declare no competing interests.

\section*{Data Availability}

The data that support the findings of this study are available from the corresponding author upon reasonable request.



\section*{Acknowledgements}

This work was primarily supported by the Center for Molecular Quantum Transduction, an Energy Frontier Research Center funded by the U.S. Department of Energy, Office of Science, Office of Basic Energy Sciences, under Award No. DE-SC0021314. Partial support was also provided by the National Science Foundation Materials Research Science and Engineering Center at Northwestern University under Award No. DMR-2308691. The research performed at the Naval Research Laboratory was supported by core programs. This research was performed while S.-J.L. held an American Society for Engineering Education fellowship at NRL.

\clearpage
\newpage

\bibliography{Citations}

\end{document}